\documentclass[prc,preprint]{revtex4}

\usepackage{graphicx}

\begin{document}

\title{Calculation of  nuclear matter in the presence of strong magnetic field using LOCV technique}

\author{ G. H. Bordbar and Z. Rezaei}
 \affiliation{Physics Department, Shiraz University, Shiraz 71454, Iran\\
and\\
Center for Excellence in Astronomy and Astrophysics (CEAA-RIAAM)-Maragha, P.O. Box 55134-441, Maragha 55177-36698, Iran}


\begin{abstract}
In the present work, we are interested in the properties of nuclear matter at zero temperature in the presence of strong magnetic fields using the lowest order constraint variational (LOCV) method employing $AV_{18}$ nuclear potential. Our results indicate that in the absence of a magnetic field, the energy per particle is a symmetric function of the spin polarization parameter. This shows that for the nuclear matter, the spontaneous phase transition to a ferromagnetic state does not occur. However, we have found that for the magnetic fields $ B\gtrsim 10 ^ {18}\ G$, the symmetry of energy is broken and the energy has a minimum at a positive value of the spin polarization parameter. We have also found that the effect of magnetic field on the value of energy is more significant at the low densities. Our calculations show that at lower densities, the spin polarization parameter is more sensitive to the magnetic field.
\end{abstract}

\maketitle



\section{INTRODUCTION}

New discoveries related to the magnetic field of a neutron star
have led to the theoretical researches on the magnetic properties of
the dense matter. The
magnetic field of a neutron star may originate from the compression
of magnetic flux inherited from the progenitor star \cite{Reisen}.
Using this idea, Woltjer has predicted a magnetic field strength of
order $10^{15}\ G$ for neutron stars \cite{Woltjer}. Moreover,
general relativity predicts the allowed maximum value of neutron
star magnetic field to be about $10^{18}-10^{20}\ G$ \cite{shap}. By
comparing with the observational data, Yuan et al. obtained a
magnetic field strength of order $10^{19}\ G$ for neutron stars
\cite{Yuan}. In the core of  inhomogeneous
gravitationally bound dense magnetars, the magnetic field strength can be
as large as $10^{20}\ G$ \cite{Ferrer}. In addition, considering the
formation of a quark core in the interior of a
magnetar, the maximum field reaches up to about $10^{20}\ G$
\cite{Ferrer,Tatsumi}.

These very intense magnetic fields have significant effects on
the dense nuclear matter.
Some studies have investigated the properties of neutron
star matter and nuclear matter \cite{Dong,Rabhi,Aguirre,Dong2,Haber,Vishal,Rabhi15}
in strong magnetic fields.
The instabilities of nuclear matter was studied considering the
relativistic nuclear models \cite{Rabhi}. They found that
the presence of the magnetic field will generally increase the
instability region.
%
The effects of strong magnetic fields on nuclear matter were studied in the
framework of the relativistic mean field models FSU-Gold by including the anomalous
magnetic moments of the nucleons \cite{Dong2}.
They concluded that at low densities, by increasing the magnetic field, the energy per particle turns out
to be increasing lower and a softening of the equation of state appears.
However, they showed that at high densities, while the softening of the EOS
will be gradually overwhelmed by stiffening resulting from the
anomalous magnetic moments effect, the energies are slightly reduced by a strong magnetic field.
Employing two relativistic field-theoretical
models for nuclear matter, the Walecka model and an extended linear sigma model,
it has been found that
the creation of nuclear matter in a sufficiently strong magnetic field becomes energetically more costly due
to the heaviness of magnetized nucleons \cite{Haber}.
Using the semi-empirical mass formula of Green
based on the liquid drop nuclear model, the onset of the neutron drip in high-density matter in the presence of a magnetic field has been investigated \cite{Vishal}. It has been found that for systems having only protons and electrons, in the presence of a magnetic field $ B\gtrsim 10 ^ {15}\ G$, neutronization occurs at a density that is at least an order of magnitude higher compared to that in a nonmagnetic
system.
The effect of a strong magnetic field on the proton spin polarization and
magnetic susceptibility of asymmetric nuclear matter has been studied within a relativistic mean-field approach \cite{Rabhi15}.

In our previous works, we have studied the spin polarized neutron
matter \cite{Bordbar75}, symmetric nuclear matter \cite{Bordbar76},
asymmetric nuclear matter \cite{Bordbar77}, and neutron star matter
\cite{Bordbar77} at zero temperature using the lowest order constraint variational (LOCV) method with the
realistic strong interaction in the absence of magnetic field. We
have also investigated the thermodynamic properties of these systems at
finite temperature with no magnetic field \cite{Bordbar78,
Bordbar80,Bordbar81}. Furthermore, we have
calculated the properties of spin polarized neutron matter in the
presence of strong magnetic fields at zero \cite{Bordbar044310} and finite \cite{Bordbar044311}
temperatures. Very recently, we have investigated the temperature and density dependence of
asymmetry energy of nuclear matter \cite{BordbarRJP}.
In the present work, we intend to extract the properties of
nuclear matter in strong magnetic fields by LOCV method using $AV_{18}$ potential.

\section{LOCV formalism for nuclear matter}
We study a system consists of $A$ nucleons where $A^{(+)}$ nucleons are spin-up and $A^{(-)}$ nucleons are spin-down with the spin
polarization parameter, $\xi$, as
\begin{equation}
     \xi=\frac{A^{(+)}-A^{(-)}}{A^{(+)}+A^{(-)}}.
\end{equation}
For calculation of nuclear matter properties, we apply LOCV approach. In this method, we consider up to
the two-body cluster energy \cite{Bordbar97},
\begin{equation}\label{tener1}
          E=E _{1}+E _{2}.
 \end{equation}
In above equation, $E_1$ is the total one-body energy of spin-up and spin-down nucleons,
\begin{equation}\label{tener1}
          E_1=E_1^{(+)}+E_1^{(-)}.
 \end{equation}
and $E_2$ is the two-body energy contribution,
\begin{equation}
    E_{2}=\frac{1}{2A}\sum_{ij} \langle ij\left| \nu(12)\right|
    ij-ji\rangle,
 \end{equation}
in which $\nu(12)=-\frac{\hbar^{2}}{2m}[f(12),[\nabla
_{12}^{2},f(12)]]+f(12)V(12)f(12)$ is the two-body effective potential. Here, $V(12)$ is the two-body nuclear
potential and $f(12)$ is the two-body correlation function.
To calculate the interaction energy of  the nuclear matter using LOCV formalism, the two-body correlation function, $f(12)$, is
considered as follows \cite{Owen},
\begin{equation}
f(12)=\sum^3_{k=1}f^{(k)}(r_{12})P^{(k)}_{12},
\end{equation}
where
\begin{equation}
P^{(k=1-3)}_{12}=(\frac{1}{4}-\frac{1}{4}\sigma_{1}.\sigma_{2}),\
(\frac{1}{2}+\frac{1}{6}\sigma_{1}.\sigma_{2}+\frac{1}{6}S_{12}),\
(\frac{1}{4}+\frac{1}{12}\sigma_{1}.\sigma_{2}-\frac{1}{6}S_{12}).
\end{equation}
In the above equation $S_{12}$ and $\sigma_{1}$ and $\sigma_{2}$ are the tensor and Pauli operators, respectively.
Using the above two-body correlation function and
the $AV_{18}$ two-body potential \cite{Wiringa}, we find the following equation
for the two-body energy:
\begin{eqnarray}\label{ener2}
    E_{2} &=& \frac{2}{\pi ^{4}\rho }\left( \frac{\hbar^{2}}{2m}\right)
    \sum_{JLTSS_{z}}\frac{(2J+1)(2T+1)}{2(2S+1)}[1-(-1)^{L+S+T}]\left| \left\langle
\frac{1}{2}\sigma _{z1}\frac{1}{2}\sigma _{z2}\mid
SS_{z}\right\rangle \right| ^{2}\int dr  \nonumber
\\&& \left\{\left [{f_{\alpha
}^{(1)^{^{\prime }}}}^{2}{a_{\alpha
}^{(1)}}^{2}(k_{F}r)\right.\right.\left.\left.
+\frac{2m}{\hbar^{2}}(\{V_{c}-3V_{\sigma } +(V_{\tau }-3V_{\sigma
\tau })(4T-3)-(V_{T}-3V_{\sigma T })(4T) \}\right.\right. \nonumber
\\&&\left.\left. \times{a_{\alpha
}^{(1)}}^{2}(k_{F}r)+[V_{l2}-3V_{l2\sigma } +(V_{l2\tau
}-3V_{l2\sigma \tau })(4T-3)]{c_{\alpha
}^{(1)}}^{2}(k_{F}r))(f_{\alpha }^{(1)})^{2}] \right.\right.
\nonumber
\\&&\left. \left. +\sum_{k=2,3}[
{f_{\alpha }^{(k)^{^{\prime }}}}^{2}{a_{\alpha
}^{(k)}}^{2}(k_{F}r)+\frac{2m}{\hbar^{2}}( \{V_{c}+V_{\sigma
}+(-6k+14)V_{t}-(k-1)V_{ls}\right.\right. \nonumber
\\&&\left.\left. +[V_{\tau } +V_{\sigma
\tau}+(-6k+14)V_{t\tau}-(k-1)V_{ls\tau }](4T-3)-[V_{T}+V_{\sigma T
}\right.\right. \nonumber
\\&&\left.\left. +(-6k+14)V_{tT}](4T)\}{a_{\alpha }^{(k)}}^{2}(k_{F}r)
+[V_{l2}+V_{l2\sigma } +(V_{l2\tau }+V_{l2\sigma \tau
})(4T-3)]{c_{\alpha }^{(k)}}^{2}(k_{F}r \right)\right. \nonumber
\\&&\left. +[(V_{ls2}+V_{ls2\tau })(4T-3)] {d_{\alpha
}^{(k)}}^{2}(k_{F}r)) {f_{\alpha }^{(k)}}^{2}
]+\frac{2m}{\hbar^{2}}\{[(V_{ls\tau }-2(V_{l2\sigma \tau }+V_{l2\tau
})-3V_{ls2\tau })\right. \nonumber
\\&&\left. \times(4T-3)]+V_{ls}-2(V_{l2}+V_{l2\sigma
})-3V_{ls2}\}b_{\alpha }^{2}(k_{F}r)f_{\alpha }^{(2)}f_{\alpha
}^{(3)}\right. \nonumber
\\&&\left.+\frac{1}{r^{2}}(f_{\alpha }^{(2)} -f_{\alpha
}^{(3)})^{2}b_{\alpha }^{2}(k_{F}r)\right\},
 \end{eqnarray}
where $\alpha=\{J,L,T,S,S_z\}$ and the coefficient  ${a_{\alpha
}^{(1)}}^{2}$, etc., are defined as
\begin{equation}\label{a1}
     {a_{\alpha }^{(1)}}^{2}(x)=x^{2}I_{L,S_{z}}(x),
 \end{equation}
\begin{equation}
     {a_{\alpha }^{(2)}}^{2}(x)=x^{2}[\beta I_{J-1,S_{z}}(x)
     +\gamma I_{J+1,S_{z}}(x)],
 \end{equation}
\begin{equation}
           {a_{\alpha }^{(3)}}^{2}(x)=x^{2}[\gamma I_{J-1,S_{z}}(x)
           +\beta I_{J+1,S_{z}}(x)],
      \end{equation}
\begin{equation}
     b_{\alpha }^{(2)}(x)=x^{2}[\beta _{23}I_{J-1,S_{z}}(x)
     -\beta _{23}I_{J+1,S_{z}}(x)],
 \end{equation}
\begin{equation}
         {c_{\alpha }^{(1)}}^{2}(x)=x^{2}\nu _{1}I_{L,S_{z}}(x),
      \end{equation}
\begin{equation}
        {c_{\alpha }^{(2)}}^{2}(x)=x^{2}[\eta _{2}I_{J-1,S_{z}}(x)
        +\nu _{2}I_{J+1,S_{z}}(x)],
 \end{equation}
\begin{equation}
       {c_{\alpha }^{(3)}}^{2}(x)=x^{2}[\eta _{3}I_{J-1,S_{z}}(x)
       +\nu _{3}I_{J+1,S_{z}}(x)],
 \end{equation}
\begin{equation}
     {d_{\alpha }^{(2)}}^{2}(x)=x^{2}[\xi _{2}I_{J-1,S_{z}}(x)
     +\lambda _{2}I_{J+1,S_{z}}(x)],
 \end{equation}
\begin{equation}\label{d2}
     {d_{\alpha }^{(3)}}^{2}(x)=x^{2}[\xi _{3}I_{J-1,S_{z}}(x)
     +\lambda _{3}I_{J+1,S_{z}}(x)],
 \end{equation}
with
\begin{equation}
          \beta =\frac{J+1}{2J+1},\ \gamma =\frac{J}{2J+1},\
          \beta _{23}=\frac{2J(J+1)}{2J+1},
 \end{equation}
\begin{equation}
       \nu _{1}=L(L+1),\ \nu _{2}=\frac{J^{2}(J+1)}{2J+1},\
       \nu _{3}=\frac{J^{3}+2J^{2}+3J+2}{2J+1},
      \end{equation}
\begin{equation}
     \eta _{2}=\frac{J(J^{2}+2J+1)}{2J+1},\ \eta _{3}=
     \frac{J(J^{2}+J+2)}{2J+1},
 \end{equation}
\begin{equation}
     \xi _{2}=\frac{J^{3}+2J^{2}+2J+1}{2J+1},\
     \xi _{3}=\frac{J(J^{2}+J+4)}{2J+1},
 \end{equation}
\begin{equation}
     \lambda _{2}=\frac{J(J^{2}+J+1)}{2J+1},\
     \lambda _{3}=\frac{J^{3}+2J^{2}+5J+4}{2J+1},
 \end{equation}
and
\begin{equation}
       I_{J,S_{z}}(x)=\int dq\ q^2 P_{S_{z}}(q) J_{J}^{2}(xq)\cdot
 \end{equation}
$J_{J}(x)$ is the familiar Bessel function and
$P_{S_{z}}(q)$ is defined as
\begin{eqnarray}
P_{S_{z}}(q)&=&\frac{2}{3}\pi[(k_{F} ^{\sigma_{z1}})^{3}+(k_{F}
^{\sigma_{z2}})^{3}-\frac{3}{2}((k_{F} ^{\sigma_{z1}})^{2}+(k_{F}
^{\sigma_{z2}})^{2})q\nonumber\\ &-&\frac{3}{16}((k_{F}
^{\sigma_{z1}})^{2}-(k_{F} ^{\sigma_{z2}})^{2})^{2}q^{-1}+q^{3}]
 \end{eqnarray}
for $\frac{1}{2}|k_{F} ^{\sigma_{z1}}-k_{F}
^{\sigma_{z2}}|<q<\frac{1}{2}|k_{F} ^{\sigma_{z1}}+k_{F}
 ^{\sigma_{z2}}|$,
\begin{equation}
P_{S_{z}}(q)=\frac{4}{3}\pi min((k_{F} ^{\sigma_{z1}})^3,(k_{F}
^{\sigma_{z2}})^3)
 \end{equation}
for  $q<\frac{1}{2}|k_{F} ^{\sigma_{z1}}-k_{F}
 ^{\sigma_{z2}}|$, and
 \begin{equation}
       P_{S_{z}}(q)=0
 \end{equation}
for  $q>\frac{1}{2}|k_{F} ^{\sigma_{z1}}+k_{F}
 ^{\sigma_{z2}}|$, where $k_{F}^{(i)}=(3\pi^{2}\rho^{(i)})^{\frac{1}{3}}$ and $\sigma_{z1}$ or
 $\sigma_{z2}= +1,-1$ for spin-up and spin-down nucleons,
 respectively.

According to LOCV formalism, the two-body energy is minimized with respect to the
variations in the functions $f_{\alpha}^{(i)}$ subject to the
normalization constraint \cite{Bordbar57},
\begin{equation}
        \frac{1}{A}\sum_{ij}\langle ij\left| h_{S_{z}}^{2}
        -f^{2}(12)\right| ij\rangle _{a}=0.
 \end{equation}
In the case of spin polarized nuclear matter, the function
$h_{S_{z}}(r)$ is introduced as follows,

\begin{eqnarray}
h_{S_{z}}(r)&=& \left\{\begin{array}{ll} \left[ 1-\frac{9}{2}\left(
\frac{J_{J}^{2}
(k_{F}^{(i)}r)}{k_{F}^{(i)}r}\right) ^{2}\right] ^{-1/2} &;~~ S_{z}=\pm1   \\ \\
1 &;~~ S_{z}= 0
\end{array}
\right.
\end{eqnarray}

The minimization of the two-body cluster energy leads to a set of
differential equations with the same form as
those presented in Ref. \cite{Bordbar57}, with coefficients
replaced by Eqs. (\ref{a1})$-$(\ref{d2}).
The correlation functions, and then the two-body energy term are obtained
through the solving of the differential equations.

Since, we consider the nuclear matter which is under the influence of a strong magnetic
field ($B$), we must add the contribution of magnetic energy to the energy of system (Eq. (\ref{tener1})). By considering the
magnetic field along the z axis, the contribution of the magnetic energy per nucleon is given by
\begin{equation}
E_M=-\frac{1}{A}\sum_i \overrightarrow{\mu_i}.\overrightarrow{B}=\mu\xi B,
\end{equation}
where $\mu$ is the magnetic dipole moment of nucleons.
\section{RESULTS}

\begin{figure}[h!tb]
\centering
\includegraphics{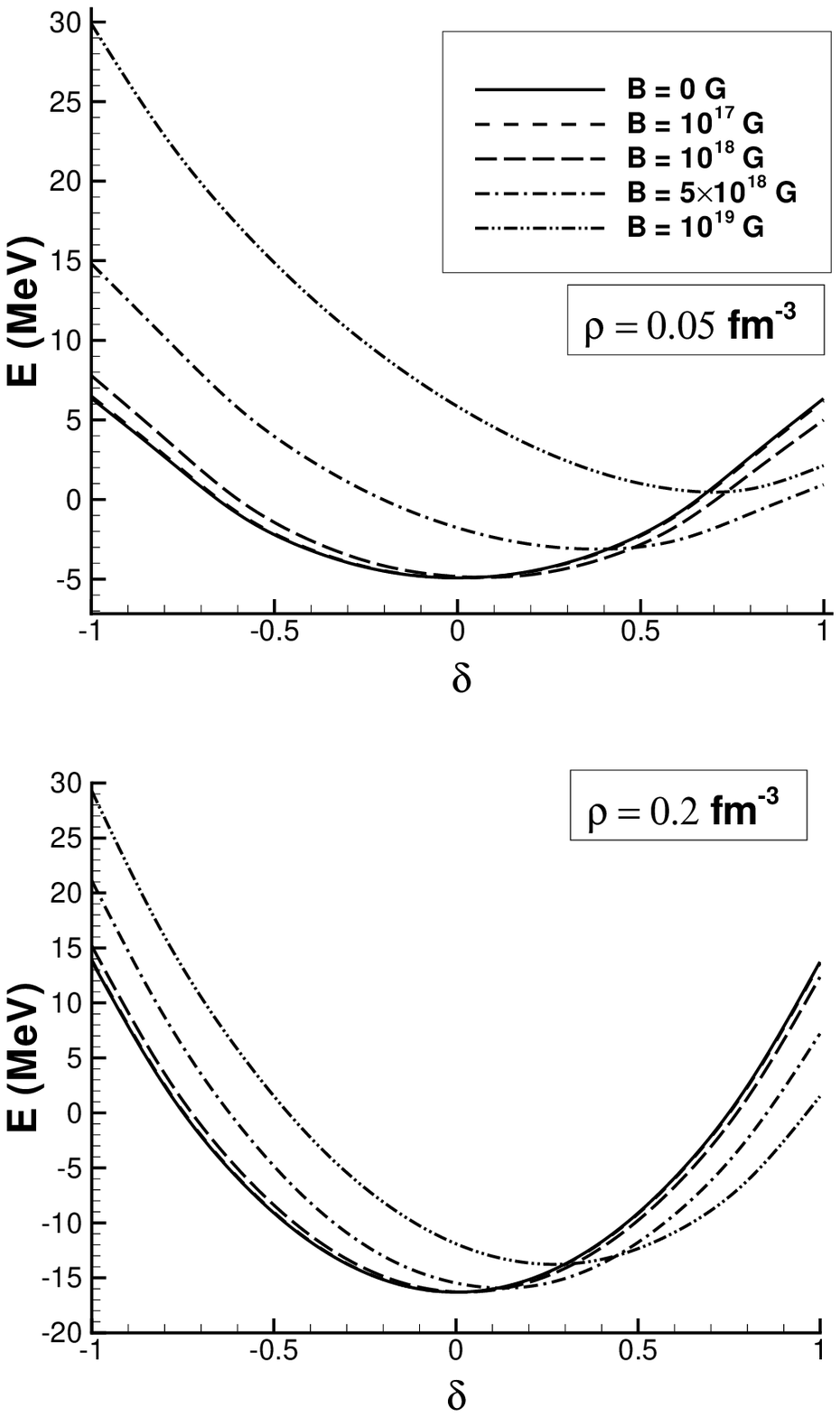}
\caption{\label{fig:1j}The energy per particle versus the spin
 polarization parameter for the cases $B=0\ G$
(solid curve), $B=10^{17}\ G$ (dashed curve), $B=10^{18}\ G$ (long dashed curve), $B=5\times10^{18}\
G$ (dashdot curve) and $B=10^{19}\
G$ (dashdotdot curve), and two values of the density, $\rho=0.05\ fm^{-3}$ (top panel) and $\rho=0.2\ fm^{-3}$ (bottom panel).}
\end{figure}

Fig. \ref{fig:1j} presents the dependence of the energy per particle on the spin polarization parameter at different magnetic fields and densities. Obviously, in the absence of a magnetic field, the energy per particle is a symmetric function of the spin polarization parameter. This shows that for the nuclear matter, the spontaneous phase transition to
a ferromagnetic state does not occur \cite{Bordbar76}.
Obviously, for $ B\gtrsim 10 ^ {18}\ G$, the spin polarization symmetry is broken and the energy has a
minimum value at a positive spin polarization parameter. The stronger the magnetic field is, the more the symmetry is broken.
The comparison between top and bottom panels shows that the effects of magnetic fields are more important at lower densities.
The degree of symmetry breaking, as well as the difference of the energy in strong magnetic fields and zero fields, is clearly higher at lower densities.
Fig. \ref{fig:1j} also shows that for $ B\gtrsim 10 ^ {18}\ G$, at each density and spin polarization parameter, the energy grows with increasing the magnetic field.

\begin{figure}[h!tb]
\centering
\includegraphics{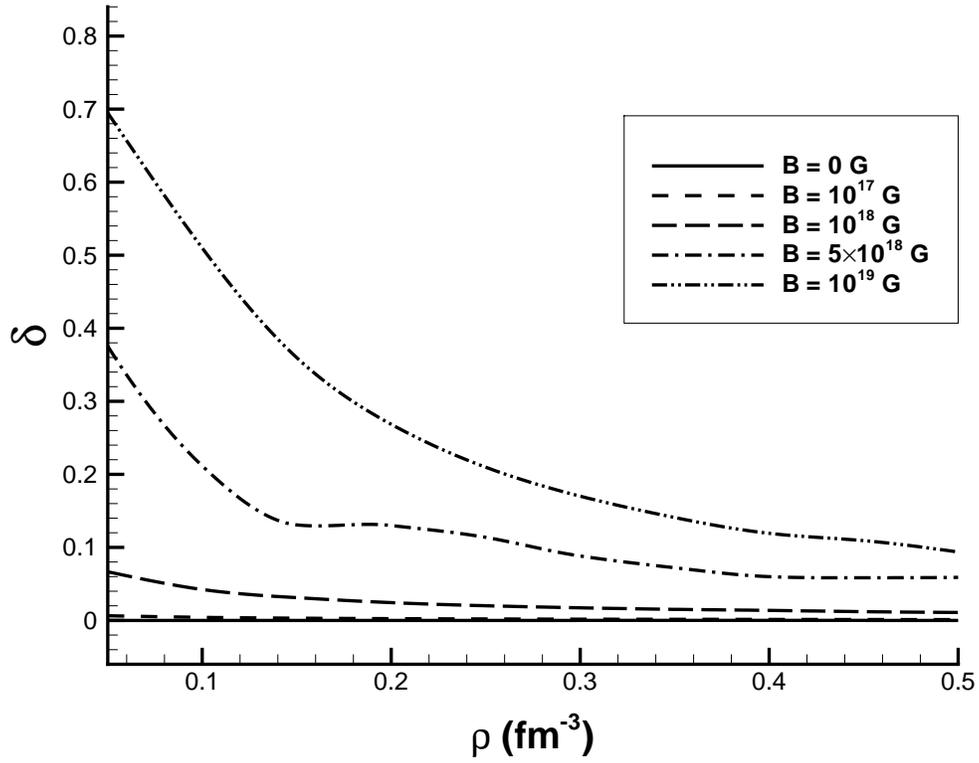}
\caption{\label{fig:2j} Spin polarization parameter at the
equilibrium state as a function of the density for the cases $B=0\ G$
(solid curve), $B=10^{17}\ G$ (dashed curve), $B=10^{18}\ G$ (long dashed curve), $B=5\times10^{18}\
G$ (dashdot curve) and $B=10^{19}\
G$ (dashdotdot curve).}
\end{figure}

The equilibrium value for the spin polarization parameter
varies with both magnetic field and density, as shown in Fig. \ref{fig:2j}.
As we mentioned above, in the free field case, the spin polarization parameter is equal to zero,
showing that no spontaneous phase transition to a ferromagnetic state occurs.
For the magnetic fields $ B \lesssim  10^{17}\ G$, the
nuclear matter is nearly unpolarized, but at
$B \gtrsim 10^{18}\ G$, the spin polarization parameter grows as the
magnetic field increases.
This parameter is more significantly affected by the magnetic field at lower densities.
For example, at $\rho =0.05\ fm^{-3}$ and $B=10 ^ {19} \ G$, the nuclear matter is polarized with $\xi\simeq0.69$.
For each value of the magnetic field, the spin polarization parameter decreases by increasing the density. This behavior has been also reported in Ref. \cite{Dong2}. Our results show that at $\rho=0.5\ fm^{-3}$ and $B=10 ^ {19} \ G$, the spin polarization parameter is less than $0.2$.

\begin{figure}[h!tb]
\centering
\includegraphics{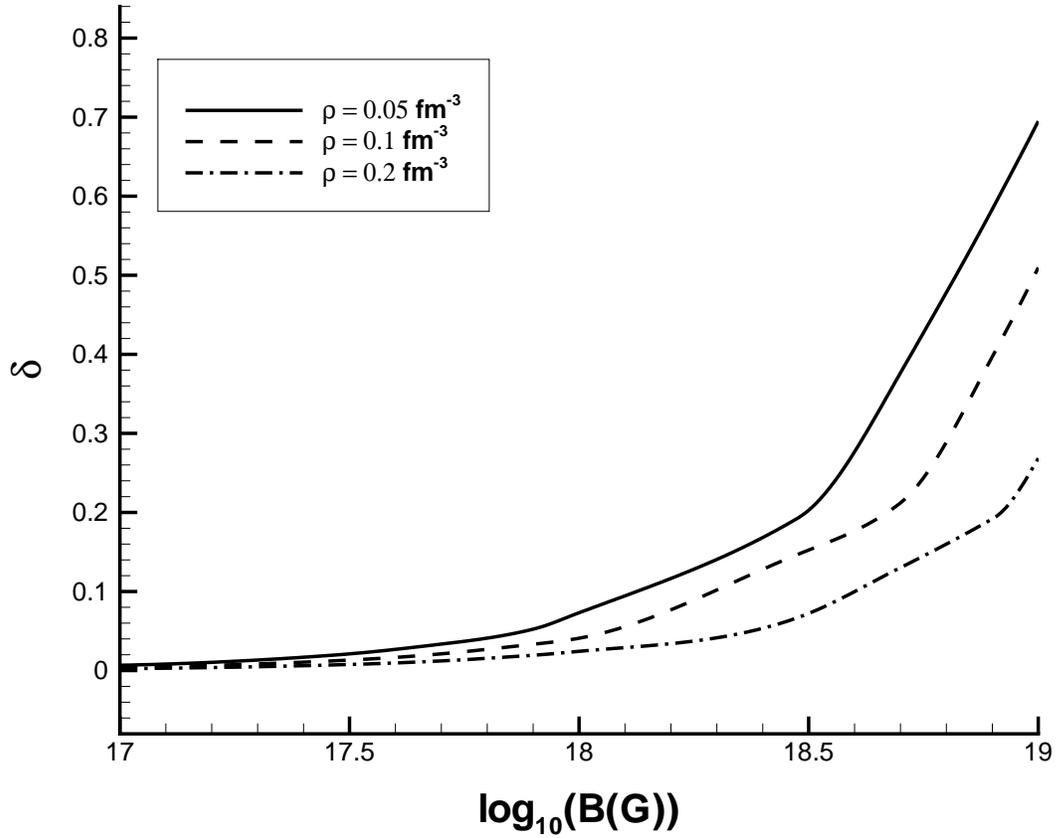}
\caption{\label{fig:3j} Spin polarization parameter at the
equilibrium state as a function of the magnetic field for the cases $\rho=0.05\ fm^{-3}$
(solid curve), $\rho=0.1\ fm^{-3}$ (dashed curve) and $\rho=0.2\ fm^{-3}$ (dashdot curve). }
\end{figure}

The magnetic field dependence of the spin polarization parameter at different densities is presented in Fig. \ref{fig:3j}.
We found that at each density, by increasing the magnetic field up to $B\simeq 10 ^ {17} \ G$, the nuclear spin polarization parameter is nearly constant.
However, for the magnetic fields $B\gtrsim 3 \times 10 ^ {17}\ G$, this parameter grows by increasing the magnetic field.
Evidently, as the density increases, the stronger magnetic fields are needed to affect the spin polarization parameter, consistent with the results obtained
in Ref. \cite{Dong2}.
The increasing rate of the spin polarization parameter versus the magnetic field depends on the nucleon density.
At lower densities, this parameter is more sensitive to the magnetic field and grows more rapidly.
Obviously, at each density, the increasing rate of the spin polarization parameter becomes significant at a particular value of the magnetic field. For example at $\rho=0.05\ fm^{-3}$, the increasing rate of $\xi$ raises at $B\simeq 3 \times 10 ^ {18}\ G$. This value of the magnetic field increases by increasing the density.

\begin{figure}[h!tb]
\centering
\includegraphics{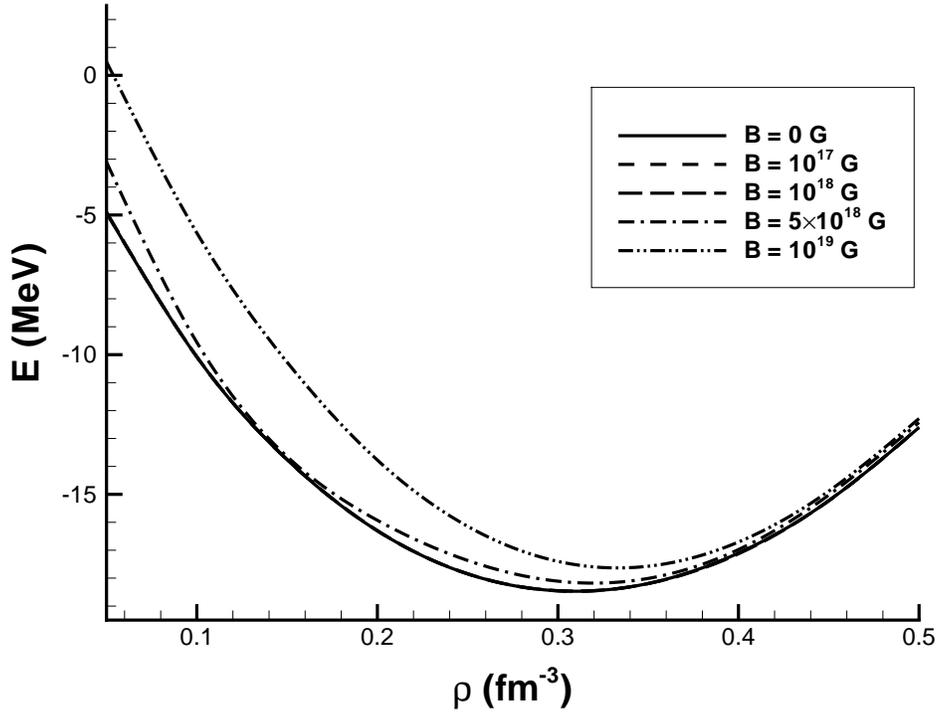}
\caption{\label{fig:4j}The energy as a function of the density for the cases $B=0\ G$
(solid curve), $B=10^{17}\ G$ (dashed curve), $B=10^{18}\ G$ (long dashed curve), $B=5\times10^{18}\
G$ (dashdot curve) and $B=10^{19}\
G$ (dashdotdot curve).}
\end{figure}

\begin{figure}[h!tb]
\centering
\includegraphics{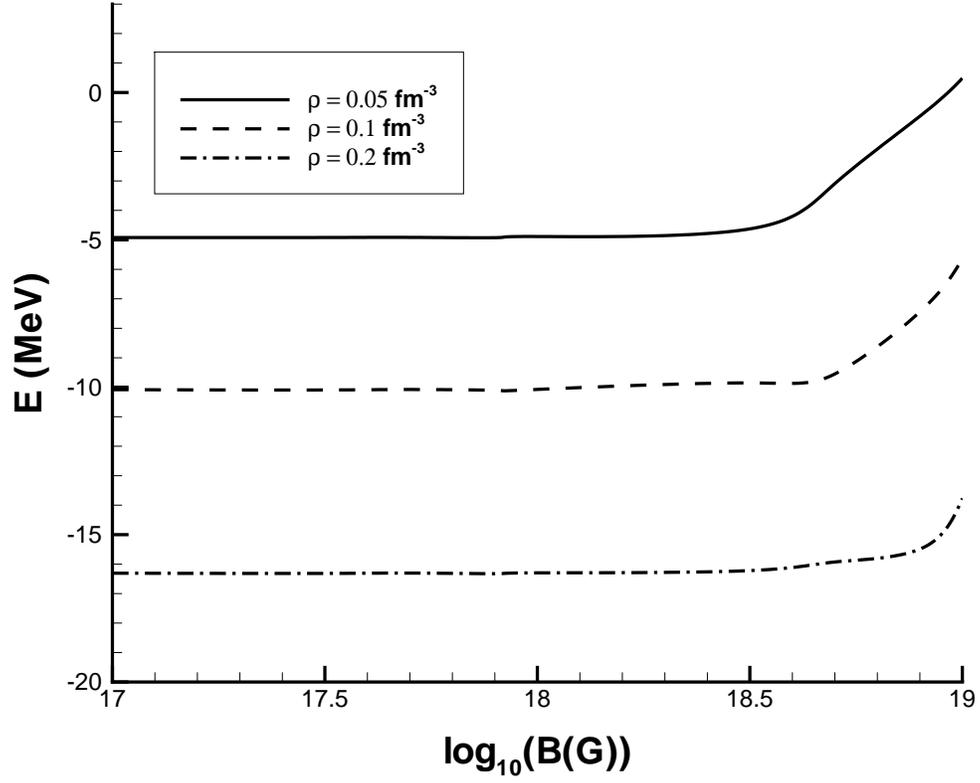}
\caption{\label{fig:5j} The energy as a function of the magnetic field for the cases $\rho=0.05\ fm^{-3}$
(solid curve), $\rho=0.1\ fm^{-3}$ (dashed curve) and $\rho=0.2\ fm^{-3}$ (dashdot curve).}
\end{figure}

The energy for nuclear matter as a function of the density at different magnetic fields is plotted in Fig. \ref{fig:4j}.
We can see that for $B\gtrsim 10 ^ {18}\ G$, at each density, the energy of the system increases as the magnetic field grows. Evidently, the effect of strong magnetic fields is more considerable at lower densities.
 Fig. \ref{fig:4j} implies that for all magnetic fields studied in this work, the nuclear matter becomes bound at a particular value of the density.
The magnetic field dependence of the nuclear matter energy at different densities is shown in Fig. \ref{fig:5j}.
We can see that at high magnetic fields, the nuclear matter energy increases as the magnetic field grows.
The magnetic field necessary to affect the nuclear matter energy depends on the nucleon density.
The higher the density becomes, the stronger the magnetic field necessary is.
In fact, the magnetic effects on the nuclear matter are significant at low densities and high magnetic fields.
Thus, to affect the high density nuclear matter, very strong magnetic field is needed.
 Fig. \ref{fig:5j} also implies that for high magnetic fields, the increase in the nuclear matter energy is more important at lower densities.

\section{Summary and Concluding Remarks}
Using LOCV method and the $AV_{18}$
two-body potential, we have considered the effects of strong magnetic
fields on the properties of nuclear matter.
It has been shown that for the nuclear matter, the spontaneous phase transition to
a ferromagnetic state does not occur, while the spin polarization symmetry of energy is broken at high magnetic fields. We have concluded that the spin polarization parameter decreases as the density increases, while it grows with increasing the magnetic field.
Our results show that the energy of the system increases as the magnetic field grows.
Moreover, It has been found that the magnetic effects on the nuclear matter are significant at low densities and high magnetic fields.

\acknowledgements{
This work has been supported financially by the Center for Excellence in
Astronomy and Astrophysics (CEAA-RIAAM). We wish to thank the Shiraz University
Research Council.}

\newpage

\newpage
\newpage
\newpage

\newpage

\end{document}